\documentclass[journal=jacsat,manuscript=article]{achemso}
\usepackage{achemso} \usepackage{float}
\author{Yuanfei Bi} \affiliation{Department of Civil and Environmental
  Engineering, George Washington University, Washington, DC 20052}
  \author{Raffaela Cabriolu} \affiliation{Department of Civil and
  Environmental Engineering, George Washington University, Washington,
  DC 20052} \author{Tianshu Li} \affiliation{Department of Civil and
  Environmental Engineering, George Washington University, Washington,
  DC 20052} \email{tsli@gwu.edu}

\title{Heterogeneous ice nucleation controlled by the coupling of
  surface crystallinity and surface hydrophilicity}

\keywords{Heterogeneous ice nucleation, Molecular Dynamics, Nucleation
  Rate, Nucleation Mechanism}

\begin{document}

\begin{abstract}

The microscopic mechanisms controlling heterogeneous ice nucleation
are complex and remain poorly understood. Although good ice nucleators
are generally believed to match ice lattice and to bind water, counter
examples are often identified. Here we show, by advanced molecular
simulations, that the heterogeneous nucleation of ice on graphitic
surface is controlled by the coupling of surface crystallinity and
surface hydrophilicity. Molecular level analysis reveals that the
crystalline graphitic lattice with an appropriate hydrophilicity may
indeed template ice basal plane by forming a strained ice layer, thus
significantly enhancing its ice nucleation efficiency. Remarkably, the
templating effect is found to transit from within the first contact
layer of water to the second as the hydrophilicity increases, yielding
an oscillating distinction between the crystalline and amorphous
graphitic surfaces in their ice nucleation efficiencies. Our study
sheds new light on the long-standing question of what constitutes a
good ice nucleator.

\end{abstract}

\section{Introduction}

In understanding and controlling the formation of ice, perhaps the
most crucial question to answer is: What makes an effective ice
nucleation center? The extensive studies on ice formation in
supercooled cloud droplets may provide some useful insight to this
question \cite{Murray:2012jk}. In clouds, the leading candidates for
heterogeneous ice nucleation centers have been identified to be
bacteria, pollen grains, mineral dust (e.g., clay), emission from
aircraft (e.g., soot), and high-molecular-weight organic compounds
(long-chain alcohols) \cite{Cantrell:2005cb}. While general
distinction exists in their chemical nature, these ice-nucleating
agents share common structural features. For example, the surfaces of
these substances may contain chemical groups capable of forming
hydrogen bond with water. As a consequence, layers of water molecules
can be hydrogen-bonded to the surfaces. The surfaces may also exhibit
the ordering patterns that resemble the structure of ice. Therefore
water layers bound to surfaces may be ice-like, providing template for
ice to nucleate.

On the basis of laboratory experiments, general criteria for a surface
to be a good ice nucleator (IN) were summarized
\cite{Pruppacher:2007tf}, and these are: (1) IN should be highly
water-insoluble; (2) IN should have similar hydrogen bonds available
at its surface; (3) IN should have an arrangement of atoms or
molecules on its surface as close as possible to that of water
molecules in some low index plane of ice. However it should be
stressed that these criteria may only be suggestive but not
predictive, and no unique correlation can be established between ice
nucleation threshold and these surface characteristics. For example,
lattice match can be a good indicator, but it cannot account for the
fact that different materials with the same lattice mismatch with ice,
e.g., long-chain alcohol \cite{PopovitzBiro:1994gq} and lead iodide
\cite{Bryant:1960jk}, exhibit different freezing
temperatures. Amorphous or poorly ordered materials such as soot
\cite{Popovicheva:2008jn} are also known to be effective ice
nucleator, despite their non-crystalline nature.  Crystalline soluble
salts such as ammonium sulphate was also found to nucleate ice
\cite{Zuberi:2001km}. In addition, recent laboratory experiment using
droplet freezing technique \cite{Atkinson:2013fe} further showed that
non-clay mineral feldspar dominate ice nucleation in mix-phase clouds.

Here we report direct computational evidence that the crystallinity
and the hydrophilicity of the carbon surface are strongly coupled to
dominate the kinetics of ice nucleation. By systematically varying the
water-carbon interaction strength over a wide range and computing the
ice nucleation rates explicitly, we observe a rich spectrum of
heterogeneous ice nucleation behaviors on both crystalline and
amorphous carbon surfaces. Remarkably, we find that only within
certain ranges of hydrophilicity does the crystallinity of the carbon
surface play an active role in nucleating ice, and within these
ranges, crystalline carbon surface is found to be significantly more
efficient than the amorphous. In other hydrophilicity, the role of
crystalline ordering becomes negligible, and no appreciable difference
is observed in the directly computed ice nucleation rates between the
crystalline and amorphous carbon surfaces. Our study demonstrates that
the commonly adopted individual criterion for good IN alone may not be
sufficient for predicting the ice nucleation capacity of a surface,
and points at ways of engineering surfaces for controlling ice
formation.

\section{Methods}

\subsection{Calculation of ice nucleation rate.} 
The direct calculation of ice nucleation rates was conducted by
employing the forward flux sampling (FFS) method
\cite{Allen:2006p640}. In FFS, the nucleation rate $R$ was obtained
through the product of initial flux rate $\dot{\Phi}_{\lambda_0}$ and
the growth probability $P(\lambda_B|\lambda_0)$ (namely,
$R=\dot{\Phi}_{\lambda_0}P(\lambda_B|\lambda_0)$), both of which can
be calculated directly by sampling the nucleation trajectories in the
parameter space defined by the order parameter $\lambda$. The $p_B$
histogram analysis in our previous study \cite{Cabriolu:2015th} has
demonstrated that a good order parameter $\lambda$ is the number of
water molecules contained in the largest ice nucleus for both
homogeneous and heterogeneous ice nucleation on graphitic surface. The
ice-like water molecule is numerically identified by the local
bond-order parameter $q_6$ with a $q_6>0.5$ \cite{Li:2011p7738}. The
initial flux rate $\dot{\Phi}_{\lambda_0}$ is obtained by $N_0/t_0V$,
where $N_0$ is the number of successful crossings to the interface
$\lambda_0$ from basin $A$, {\em i.e.}, the spontaneous formation of
ice nucleus containing $\lambda_0$ water molecules, $t_0$ is the total
time of initial sampling, and $V$ is the volume. The growth
probability $P(\lambda_B|\lambda_0)$ is computed through
$P(\lambda_B|\lambda_0)=\prod_{i=1}^{n}P(\lambda_{i}|\lambda_{i-1})$,
where $P(\lambda_{i}|\lambda_{i-1})$ is the crossing probability for
which a trajectory starts from interface $\lambda_{i-1}$ and ends on
interface $\lambda_i$. By firing a large number $M_{i-1}$ of trial
shootings from the interface $\lambda_{i-1}$ and collecting $N_i$
successful crossings at the interface $\lambda_i$, one estimates
$P(\lambda_{i}|\lambda_{i-1})=N_i/M_{i-1}$. The statistical
uncertainty of $P(\lambda_B|\lambda_0)$ consists of both the variance
of binomial distributions of $N_i$ and the landscape variance of the
configurations collected at the previous interface $\lambda_{i-1}$
\cite{Allen:2006p32}. More details of ice nucleation rate calculations
by FFS can be found in Ref. \cite{Li:2011p7738,Cabriolu:2015th}. The
computed heterogeneous ice nucleation rates for all the conditions are
listed in Table S1 in Supporting Information.

\subsection{Systems.}

Our molecular dynamics (MD) simulations were carried out using the
monotonic-water (mW) model \cite{Molinero:2009p2000}. The carbon-water
interaction is described based on the two-body term of the mW
model. To mimic a wide range of hydrophilicity of carbon surface, the
water-carbon interaction strength $\varepsilon$ is varied
\cite{Lupi:2014gj,Cox:2015wr,Cox:2015cs} between $\varepsilon_0$ and
10$\varepsilon_0$, where $\varepsilon_0=0.13$ kcal/mol is the original
water-carbon interaction strength \cite{Lupi:2014hh} that reproduces
the experimental contact angle ($86^o$) of liquid water on
graphite. The simulations include 4096 water molecules and 1008 carbon
atoms, in a nearly cubic cell combined with a periodic boundary
condition (PBC). The No$\acute{\mbox{s}}$e-Hoover thermostat was
employed to simulate the isobaric-isothermal canonical ensemble (NPT),
with a relaxation time of 1 ps and 15 ps for the temperature and
pressure, respectively. A time step of 5 fs is used throughout the
simulations.
%% To study the non-crystalline carbon surface, an amorphous
%% graphene was created by introducing the Stone-Wales (SW) defects to
%% crystalline graphene continuously by employing the Wooten-Weaire-Winer
%% (WWW) bond-switching Monte Carlo (MC) method \cite{Wooten:1985vx}. The
%% ice nucleation rates were directly computed by employing the forward
%% flux sampling method \cite{Allen:2006p640}.

\subsection{Model of amorphous graphene.}

\begin{figure}
\begin{center}
\includegraphics[width=5 in]{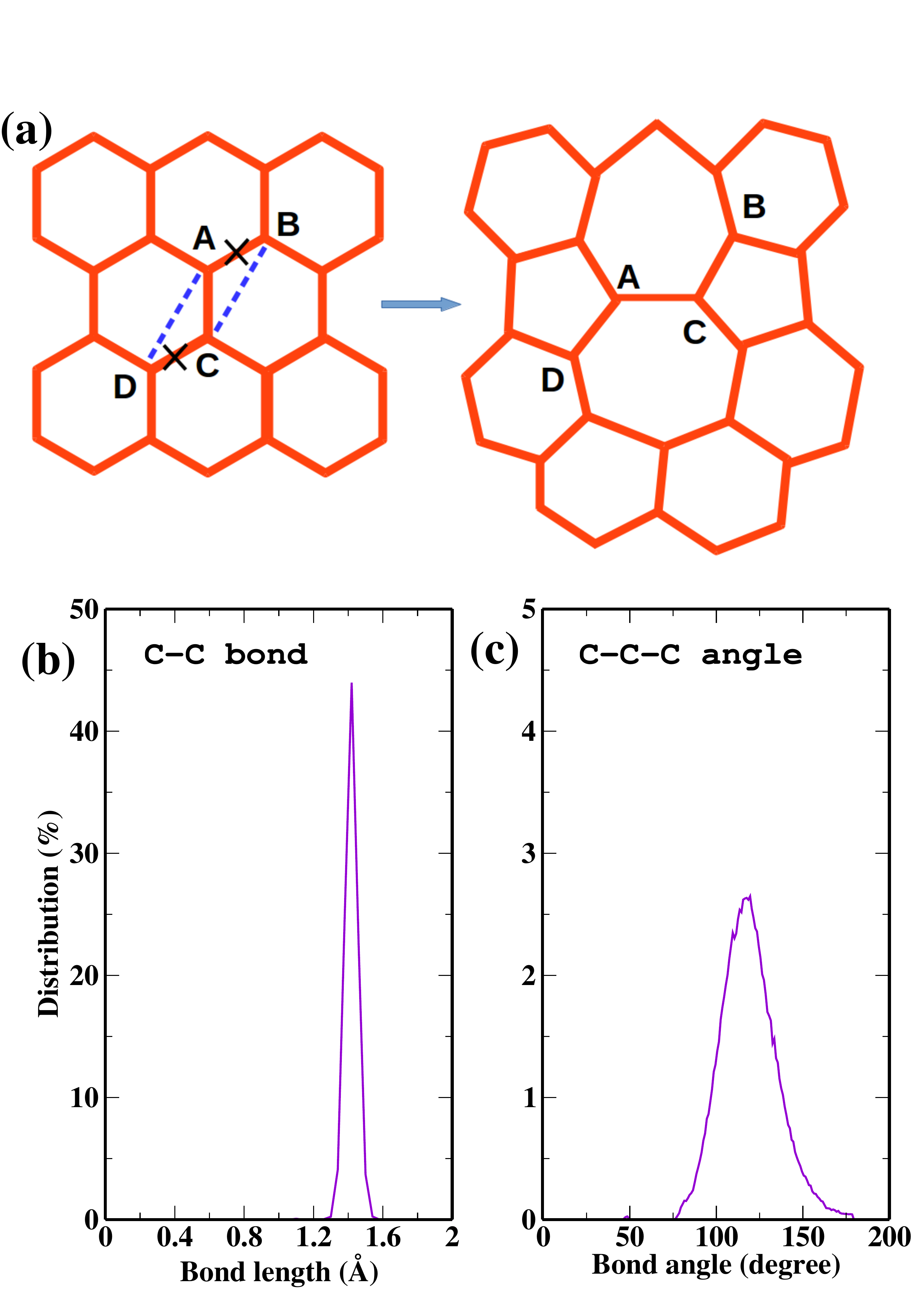}
\caption{({\bf a}) Bond interchange using the WWW algorithm to
  introduce the Stone-Wales defect in graphene. The SW defect is
  generated by rotating a carbon-carbon bond by 90 degrees. Within the
  WWW framework, this is achieved by first randomly selecting a bond,
  {\em e.g.}, AC, followed by a random selection of two carbon atoms
  connecting to A and C, respectively, but on the opposite sides of
  bond AC, {\em e.g.}, atom B and D. Then the bond AB and CD are
  replaced by new bonds AD and BC (dashed lines). The final amorphous
  graphene structure yields a distribution of the carbon-carbon bond
  length ({\bf b}) and the carbon-carbon-carbon bond angle ({\bf
  c}). }
\label{amorphous}
\end{center}
\end{figure}

To study the role of crystallinity of the carbon surface, we amorphize
the graphene surface by introducing the Stone-Wales (SW) defects
through the Wooten-Weaire-Winer (WWW) bond-switching Monte Carlo (MC)
method \cite{Wooten:1985vx}. In this approach, the carbon-carbon
interaction is described by the Keating-like potential
\cite{Keating:1966vs}:
\begin{equation} 
V=\sum_{i,j} \frac{k}{2}\left[r_{ij}^2-r_0^2 \right]^2+\sum_{i,j,k} h
\left[\vec{r}_{ki}\cdot \vec{r}_{kj}+\frac{r_0^2}{2} \right]^2 \;,
\end{equation}   
where $r_0=1.42$ \AA$\,$ which denotes the equilibrium C-C bond
length. The first and second terms of the potential describe the
energies for bond stretching and bond bending, with $k$=7.5 kcal/mol
and $h$ ($h/k=0.2$) being the corresponding spring constants,
respectively. The MC simulation includes two types of moves: a single
particle displacement and a WWW bond-switching move. The latter move
allows introducing five- and seven-fold rings in the graphene sheet
while preserving the local three fold coordination of carbon
atoms. The schematic diagram for the bond-switching move is shown in
Fig. \ref{amorphous}. By progressively introducing the bond-switching
move, one may disrupt the translational symmetry of the graphene while
minimizing the coordinational defect. In our procedure of amorphizing
graphene, each MC step consists of one attempt of single particle
displacement and one bond-switching move, both performed randomly
throughout the graphene sheet. The standard Metropolis MC procedure
was employed. To allow the structure to relax sufficiently,
particularly once the bond-switching move is accepted, the single
particle move and the bond-switching move are performed at $k_BT=0.33$
eV and 39 eV, respectively. After $10^9$ MC steps, the structure was
quenched at $k_BT=0.033$ eV for obtaining realistic atomic structure
for amorphous graphene at room temperature. Fig. \ref{amorphous}
(b)\&(c) show the distributions of the C-C bond length and C-C-C bond
angle of the amorphous graphene. The topology analysis shows the
amorphous graphene network is composed of 35\% pentagons, 37\%
hexagons, 22\% heptagons, and 4.9\% octagons, in agreement with the
ring statistics (34.5\%, 38\% 24\%, and 4.5\%, respectively) obtained
by Kapko {\em et. al.}\cite{Kapko:2010fy}.

\section{Results and Discussion}

\begin{figure}
\begin{center}
\includegraphics[width=6.6 in]{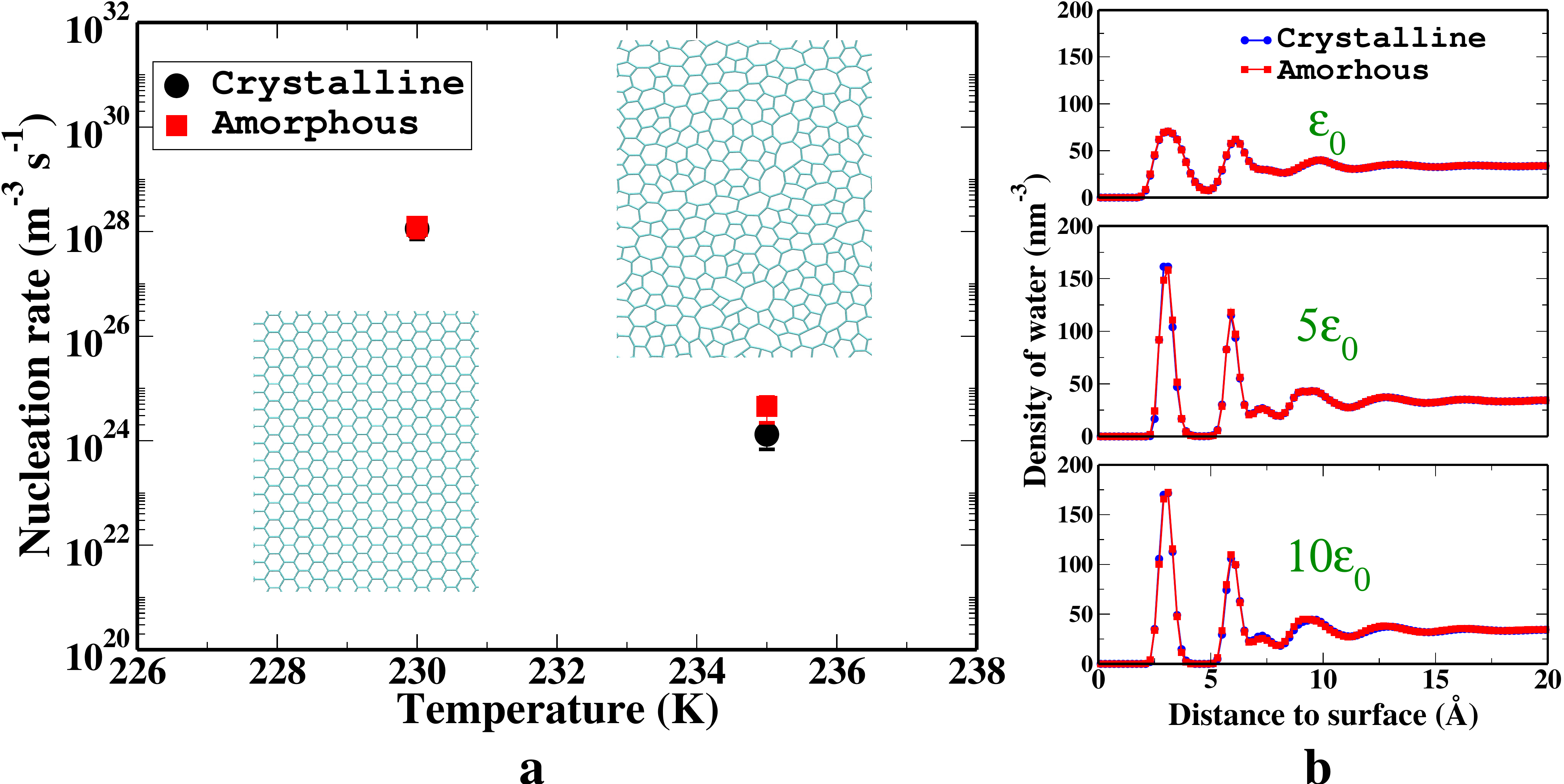}
\caption{({\bf a}) Calculated heterogeneous ice nucleation rates on
  crystalline and amorphous graphene at 230~K and 235~K. Inset shows
  the atomic structures of both crystalline and amorphous
  graphene. ({\bf b}) Calculated density profiles of water along the
  direction normal to the water-graphene interface, for different
  water-carbon interaction strengths.}
\label{1e}
\end{center}
\end{figure}

We first carried out study of heterogeneous ice nucleation on carbon
surface with the original water-carbon interaction strength
($\varepsilon=\varepsilon_0$). Figure \ref{1e}a shows the computed
heterogeneous ice nucleation rates on both crystalline and amorphous
graphene. At both temperatures investigated, no distinguishable
difference is found on the computed ice nucleation rate between the
two surfaces. The result may appear surprising in its first glimpse
but can be understood by the fact that crystalline graphene neither
binds water nor templates ice. In fact the heterogeneous ice
nucleation of graphene was recently attributed to the induced layering
of water density perpendicular to water-carbon interface
\cite{Lupi:2014hh,Lupi:2014gj,Cox:2015cs}: Since the oscillation of
water density near the water-carbon interface matches well the density
profile of ice, and the motion of water molecules in the contact
layers is well restricted within the plane, the space that these water
molecules can explore, hence the entropic barrier of ice nucleation,
is effectively reduced. It has been further suggested
\cite{Cox:2015cs} that water molecules in the first layer experience a
nearly uniform potential from graphene. In other words, water
molecules in contact layers see carbon surface as an atom-less flat
surface. Indeed, our calculated density of water (Fig. \ref{1e}b )
shows that the absence of crystallinity in graphene does not alter the
profiles of water density normal to the surface. This is consistent
with the observed insensitivity of ice nucleation rate on surface
crystallinity under the original surface hydrophilicity.

\begin{figure}
\begin{center}
\includegraphics[width=6.5 in]{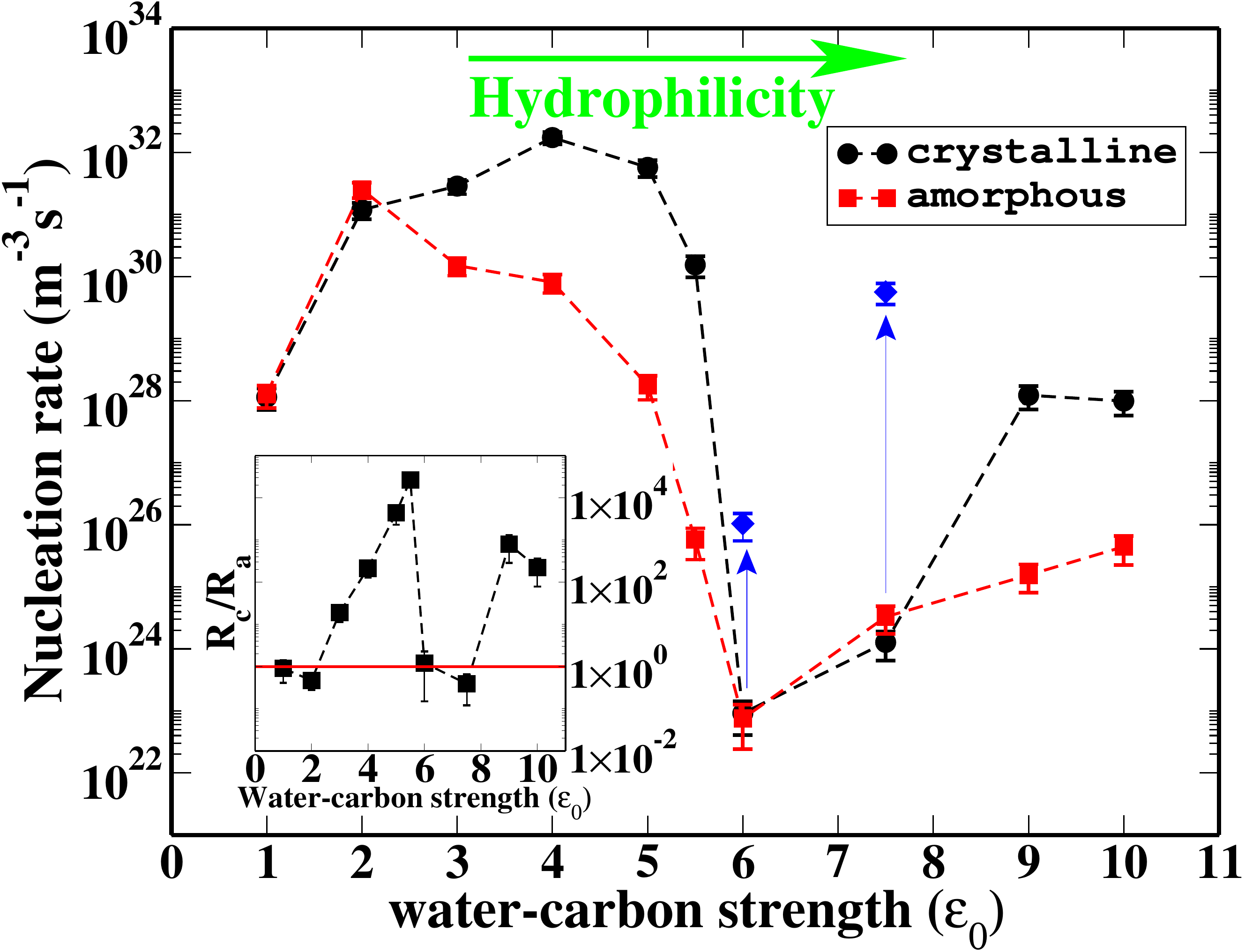}
\caption{Variation of the calculated ice nucleation rates with
  water-carbon interaction strength $\varepsilon$ (in the unit of the
  original strength $\varepsilon_0$) at 230~K, for both crystalline
  and amorphous graphene. The blue diamonds indicate the calculated
  nucleation rate of ice forming on crystalline graphene with a
  stretched carbon-carbon bond length of 1.46 \AA$\,$. Inset shows the
  ratio of the ice nucleation rates between the crystalline and
  amorphous graphene $R_c/R_a$, as a function of water-carbon
  strength. The red horizontal line indicates a ratio of one.}
\label{rate-hydro}
\end{center}
\end{figure}

The interesting results are obtained when water-carbon interaction
strength $\varepsilon$ is allowed to vary. Increasing $\varepsilon$
makes carbon atoms bind water more strongly, and correspondingly the
surface becomes more hydrophilic. As shown in Figure \ref{rate-hydro},
the variation of $\varepsilon$ (hence hydrophilicity) yields a rich
spectrum of heterogeneous ice nucleation behaviors on both crystalline
and amorphous graphene. First, increasing surface hydrophilicity
initially enhances the ice nucleation rates on both surfaces, which
maximizes the ice nucleation efficiencies for the crystalline and
amorphous at the water-carbon strength of $4\varepsilon_0$ and
$2\varepsilon_0$, respectively. A further increase of hydrophilicity,
nevertheless, starts weakening the ice nucleation capacities of both
surfaces, until the minimum nucleation rates are reached at
$6\varepsilon_0$ for both surfaces. Upon additional enhancement of
hydrophilicity, both surfaces are found to re-gain their abilities of
nucleating ice.

Remarkably, in contrast to the behaviors observed at the original
water-carbon strength $\varepsilon_0$, the crystallinity of the
graphene surface was found to play a significant role in facilitating
ice nucleation when surface hydrophilicity is varied. As shown in
Figure \ref{rate-hydro}, there exist ranges of hydrophilicity where
the crystalline graphene yields significantly higher ice nucleation
rate than the amorphous graphene. This can be better illustrated by
the inset of Fig. \ref{rate-hydro} which shows the ratio of the ice
nucleation rate resulted from the crystalline graphene to that of the
amorphous surface, $R_c/R_a$, as a function of water-carbon
interaction strength $\varepsilon$. It is evident that within the low
($\varepsilon\le 2\varepsilon_0$) and mid-high ($6\varepsilon_0 \le
\varepsilon \le 7.5 \varepsilon_0$) ranges of water-carbon interaction
strength, the crystalline graphene shows no appreciable difference
from the amorphous in its ice nucleation ability, whereas it becomes
much more efficient within other ranges. In particular, at
$\varepsilon=5\varepsilon_0$ the crystalline graphene yields an ice
nucleation rate almost $10^5$ times higher than the amorphous,
strongly suggesting that the crystallinity is a key factor for ice
nucleation under such hydrophilicity.

The results immediately suggest the existence of a coupling resulted
from the two surface characteristics that dictates ice nucleation, and
that the coupling strength varies non-monotonically with the surface
hydrophilicity.  To understand the origin of this coupling behavior,
we first examine the recently proposed layering
mechanism. Fig. \ref{1e}b shows, however, that the density profile of
water normal to the carbon-water surface is essentially insensitive to
the change of surface crystallinity, within the entire range of
hydrophilicity investigated in this work. Clearly, the observed
difference of ice nucleation rates induced by crystallinity may not be
explained by the layering mechanism.

\begin{figure*}
\begin{center}
\includegraphics[width=6.5 in]{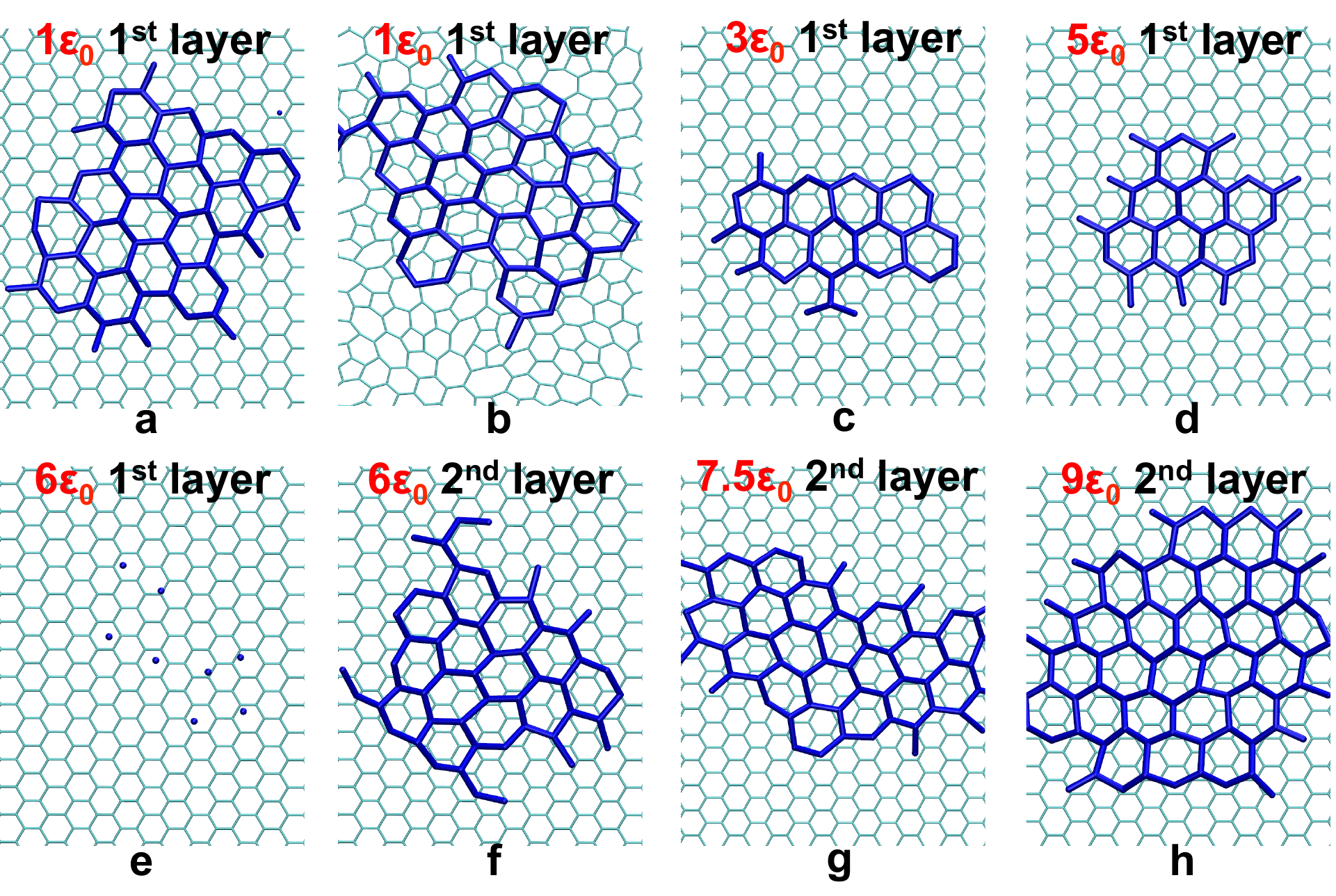}
\caption{Evolution of ice layer (blue) in the critical nucleus forming
on graphene (cyan) with water-carbon interaction strength
$\varepsilon$. ({\bf a}) and ({\bf b}) show the ice layer forming in
the first contact layer of water on the crystalline and the amorphous
graphene, respectively, at the original strength $\varepsilon_0$. With
an increasing $\varepsilon$, the ice layer gains lattice registry with
respect to the crystalline graphene, as in ({\bf c}) and ({\bf
d}). Further hydrophilicity increase disfavors ice formation in the
first contact layer, as in ({\bf e}) where only dangling water
molecules appear, and instead, facilitates ice nucleation in the
second contact layer, as in ({\bf f}). The ice layer gradually aligns
registered with the crystalline graphene with the continuous increase
of water-carbon strength, as in ({\bf h}).}
\label{icelayer}
\end{center}
\end{figure*}

To shed light on its origin, we examine the structure of the critical
ice nucleus forming on carbon surface. As heterogeneous ice nucleation
aligns the basal plane of ice parallel to graphene, the ice structure
in the contact layers are expected to play an active role in ice
nucleation. Fig. \ref{icelayer}a \& b show the structure of the first
ice layer of a critical nucleus formed on the underlying substrate
with the original $\varepsilon=\varepsilon_0$, for both crystalline
and amorphous graphene. As expected, the basal plane of ice does not
appear to match the underlying crystalline graphene
lattice. Correspondingly the computed ice nucleation rates do not
exhibit fundamental difference when the crystallinity of graphene
changes. As the water-carbon strength increases to $3\varepsilon_0$,
the first layer of ice and the graphene lattice are found to form a
nearly commensurate structure, {\em i.e.}, each six member ring of
water encloses a six member ring of carbon, as shown in
Fig. \ref{icelayer}c. On the basis of the underlying crystalline
graphene lattice, the pseudo commensurate structure can be considered
as a $3\times 3$ supercell, containing three graphene unit cells along
each direction. We note that at this hydrophilicity, the crystalline
graphene already yields an ice nucleation rate about ten times higher
than that of the amorphous graphene. As shown in Fig. \ref{icelayer}d,
the pseudo commensurate structure becomes even more evident at
$5\varepsilon_0$, but interestingly, disappears upon a further
increase of water-carbon strength to $6\varepsilon_0$. Under this
hydrophilicity ($6\varepsilon_0$), it is found that the water
molecules in the first contact ice layer no longer arrange themselves
into an ice like structure (Fig. \ref{icelayer}e). Instead, the basal
plane of ice forms in the second layer of water, and it is further
observed that the second ice layer, similar to the {\em first} ice
layer forming at the original water-carbon strength $\varepsilon_0$,
is incommensurate with the underlying graphene lattice
(Fig. \ref{icelayer}f).  We also note that at this water-carbon
strength ($6\varepsilon_0$), the computed ice nucleation rates for
both crystalline and amorphous graphene become equivalent again, and
reach their minima (see Fig. \ref{rate-hydro}). More interestingly,
upon a further increase in hydrophilicity, {\em e.g.},
$\varepsilon=9\varepsilon_0$, the basal plane of ice in the second
layer appears to form the $3\times 3$ pseudo commensurate structure,
similar to the first ice layer formed on the crystalline graphene at
$3\varepsilon_0 \sim 5\varepsilon_0$. Through the formation of the
pseudo commensurate structure, the crystalline graphene now regains
its efficiency for promoting ice nucleation.

\begin{figure*}
\begin{center}
\includegraphics[width=6.5 in]{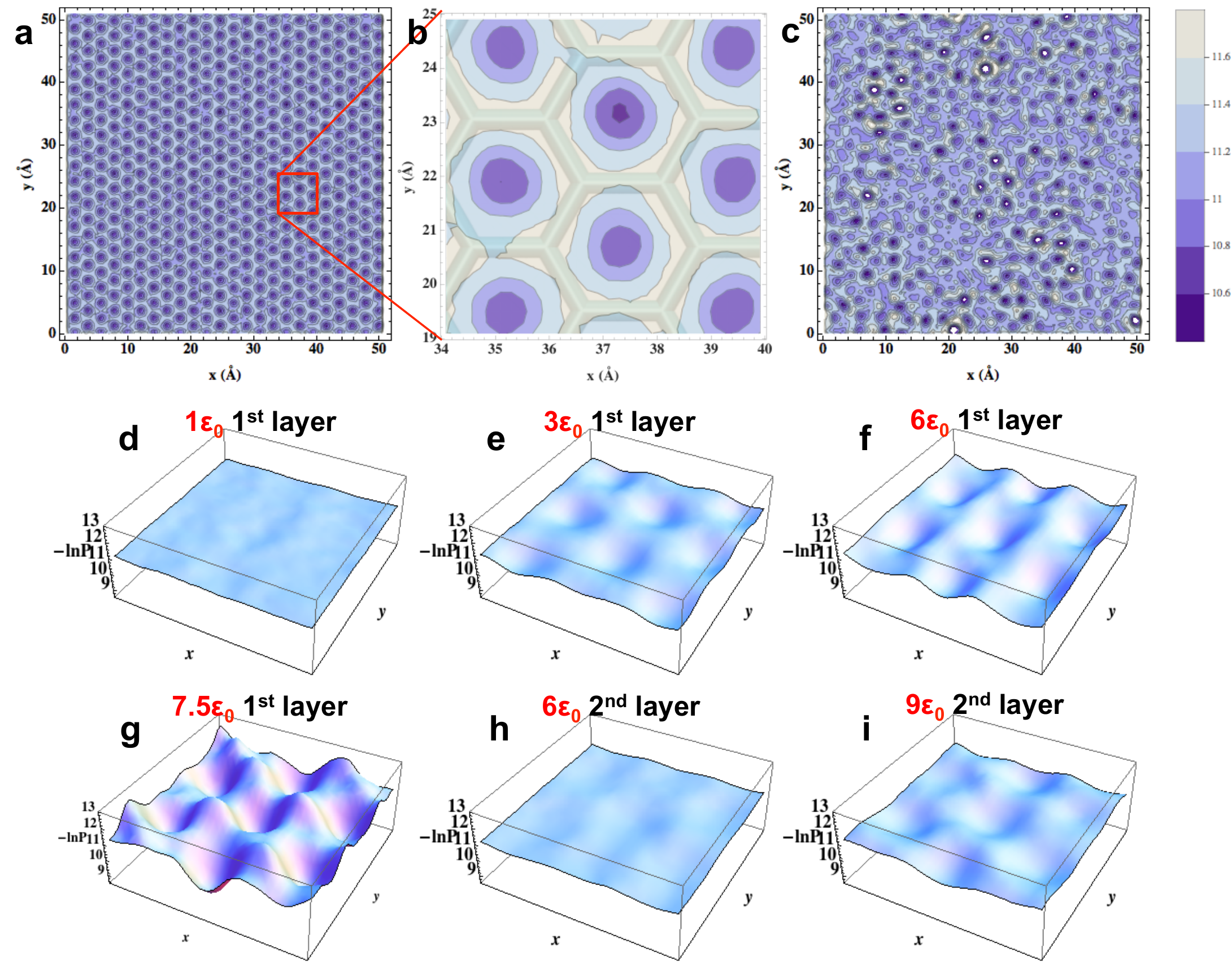}
\caption{The in-plane distribution of water molecules in the contact
  layers of graphene at 230~K. For clarity, the distribution is
  defined as $-\mbox{Ln}\left[ P(x,y)\right]$ \cite{Cox:2015cs}, where
  $P(x,y)$ is the probability density of finding a water molecule
  located at ($x$,$y$) in the plane. ({\bf a}) and ({\bf c}) show the
  contour of $-\mbox{Ln}\left[ P(x,y)\right]$ computed for the first
  water layers in contact with crystalline graphene and amorphous
  graphene, respectively, with a carbon-water strength of
  $\varepsilon=3\varepsilon_0$. As indicated by the contour legend, a
  darker color represents a higher probability of distribution. ({\bf
  b}) is the zoom-in of a local region in ({\bf a}), which shows water
  tends to be adsorbed above the center of carbon hexagonal ring. To
  demonstrate the influence of water-carbon interaction strength
  $\varepsilon$ on the distribution of water in contact layers, ({\bf
  d})$\sim$({\bf i}) show the distribution $-\mbox{Ln}\left[
  P(x,y)\right]$ as a function of both $x$ and $y$ for the same region
  as in ({\bf b}), with different hydrophilicity. In general, the
  increasing hydrophilicity of crystalline graphene patterns the first
  layer of water, which resembles the underlying graphene
  lattice. When water-carbon strength is high enough (as in ({\bf
  i})), the second layer of water is also patterned, as a result of
  the strong localization of water in the adsorption sites in the
  first layer and the local tetrahedral ordering of water. }
\label{xydensity}
\end{center}
\end{figure*}

The molecular analysis of ice nucleus suggests that the higher ice
nucleation efficiency of the crystalline graphene within
$2\varepsilon_0 \le \varepsilon \le 6 \varepsilon_0$ and
$\varepsilon\ge 7.5\varepsilon_0$ is closely related to its ability of
forming an ice layer nearly commensurate with the underlying graphene
lattice. Essentially this can be understood on the basis of the
templating effect, albeit that the templating of ice occurs over the
supercell of graphene lattice. The question now is: why does this
templating effect only occur at certain hydrophilicity?

To answer this question, we examine the in-plane density of water in
the contact layers of graphene. At a low water-carbon strength, {\em
i.e.}, when carbon atom binds water weakly, water molecules in the
first contact layer only experiences a nearly uniform potential from
the underlying graphene. This can be illustrated by the in-plane
distribution of water density in Fig. \ref{xydensity}d. When carbon
binds water more strongly ($\varepsilon\ge 3\varepsilon_0$), the
in-plane density distribution of water in the first layer begins
affected by the underlying atomic structures. In the case of
crystalline graphene, water molecules in the contact layer are found
to preferentially locate themselves above the center of the carbon
hexagonal ring (see Fig. \ref{xydensity}b), making those weak
adsorption sites of water.  Consequently, the density of water clearly
displays a pattern reminiscent to graphene lattice, as shown in
Fig. \ref{xydensity}a. The ordering of water density in the first
layer, which partially matches the ice basal plane of ice, thus may
effectively increase the possibility of forming an ice-like
fragment. In the case of amorphous graphene, the disordered, less
uniform water density in the first layer (Fig. \ref{xydensity}c) in
fact frustrates the crystalline ordering required for ice nucleation,
which leads to a decrease of ice nucleation rate. As a consequence,
the difference in the ice nucleation efficiency between a crystalline
and amorphous graphene starts increasing with hydrophilicity.

As water-carbon interaction becomes even stronger ({\em e.g.},
$6\varepsilon_0$), the water molecules in the first layer are further
constrained around the adsorption sites (Fig.\ref{xydensity}f). This
leads to two effects that both suppress ice nucleation on crystalline
graphene surface. First, because the formation of the hexagonal ice
patchworks on crystalline graphene requires that only two thirds of
the adsorption sites are filled while the rest one third are empty
(see Fig. \ref{icelayer}c \& d), a full coverage of water with strong
binding strength in fact structurally hinders ice
nucleation\cite{Cox:2015cs}.  Second, since the underlying graphene
structure is rigid with a fixed carbon-carbon bond length $d_{CC}=$1.4
\AA, the formation of an ideal commensurate superstructure that
matches both ice and graphene would require an in-plane lattice
constant of ice of 4.2 \AA, {\em i.e.}, three times of $d_{CC}$. This
implies that the ice layer in such structure must be strained
(compressed in this case), given that the equilibrium in-plane lattice
constant of ice is 4.5 \AA.  Therefore the strain energy cost of
imposing such perfect match eventually makes the formation of the
strained ice layer energetically unfavorable in the contact layer. To
further demonstrate the role of strain, we increase the carbon-carbon
bond length in crystalline graphene to 1.46 \AA$\,$, in order to
better match ice lattice. The calculated ice nucleation rate at both
$6\varepsilon_0$ and $7.5\varepsilon_0$ indeed show that the elongated
crystalline graphene yields significant enhancement on ice nucleation
rate (Fig. \ref{rate-hydro}) relative to those of the unstrained
graphene.

As the first water layer becomes inactive, the nucleation of ice
consequently starts occurring in the second layer of water when
$\varepsilon\ge 6\varepsilon_0$. In this case, as the in-plane water
density in the second layer is nearly uniform (Fig. \ref{xydensity}h),
the crystallinity of the underlying graphene becomes inactive again in
ice nucleation, as evidenced by the synchronization of the computed
ice nucleation rates within $6\varepsilon_0 \le \varepsilon \le
7.5\varepsilon_0$. Interestingly, a further increase of hydrophilicity
($\varepsilon\ge 9\varepsilon_0$) is found to nearly immobilize water
molecules within the first water layer in their adsorption sites. The
strong localization of water in the first water layer essentially
turns it into an image of the underlying crystalline graphene
sheet. The second layer of water, now under the influence of the
patterning from the first layer and the water-water interaction,
becomes also structured (Fig. \ref{xydensity}i) and tends to
facilitate the formation of the commensurate ice-like structure. In
other words, the first and the second layers of water at the high
hydrophilicity act almost as the graphene substrate and the first
layer of water at the low hydrophilicity, respectively. In this way,
the crystalline graphene sheet gains renewed ice nucleation
capability, making it again superior than amorphous graphene for
nucleating ice.

%\section{Conclusions}

The findings in this work have a few important implications regarding
our understanding of heterogeneous ice nucleation. Perhaps the most
immediate implication is that neither surface crystallinity nor
surface hydrophilicity alone may be a good indicator for the ice
nucleation efficiency of a surface. Instead, the combined surface
characteristics may yield a complex coupling that dominates ice
nucleation behaviors.  This may potentially explain why materials that
have similar lattice mismatch with ice exhibit drastically different
ice nucleation behaviors, and that no correlation has been established
between ice nucleation threshold and any of the crystallographic
characteristics \cite{Pruppacher:2007tf}. Therefore a thorough
understanding of the ice nucleation capacity for an IN should be
achieved through a comprehensive study by explicitly considering all
the necessary molecular details at the surface. Second, it is
envisioned that the surface chemistry and surface crystallinity may
also be coupled with the elasticity of the substrate. In our modeling
the graphene surface is considered rigid. This is a good approximation
as carbon-carbon bond is much stronger than the hydrogen bond of
ice. When ice nucleates on a soft substrate that has a shear modulus
lower than or comparable to that of ice, the possible local
deformation of the substrate should play an active role if there
exists a lattice mismatch between ice and the surface. Since a soft
substrate may better template ice for a lower strain energy cost, the
range of its lattice mismatch can be wider than a stiff substrate for
achieving the comparable ice nucleation efficiency. This also may
explain why some soft materials, {\em e.g.}, self-assembled monolayers
of amphiphilic alcohols \cite{PopovitzBiro:1994gq}, are known as the
most effective INs \cite{Murray:2012jk}.

\section{Acknowledgment}

The work is supported by NSF through award CMMI-1537286. T. L. also
thanks the Sloan Foundation through the Deep Carbon Observatory for
supporting this work.

%\bibliography{../../../bibtex}

\providecommand{\latin}[1]{#1}
\providecommand*\mcitethebibliography{\thebibliography}
\csname @ifundefined\endcsname{endmcitethebibliography}
  {\let\endmcitethebibliography\endthebibliography}{}

\pagebreak

%% \begin{figure}[H]
%%   \includegraphics[width=4. in]{../plot/FTOC.eps}
%%   \caption*{For Table of Contents Only}
%% \end{figure}

\end{document}